\documentclass[aps, prl, 10pt, superscriptaddress, twocolumn]{revtex4-1}
	\usepackage[bookmarks=false,linkcolor=blue,urlcolor=blue,colorlinks,citecolor=blue]{hyperref}

	\usepackage{graphicx}
	\usepackage{color}
	\usepackage{amsmath}
	\usepackage[normalem]{ulem}
	\usepackage[toc,page]{appendix}
	\usepackage{amsfonts}
	\usepackage{amssymb}
    \usepackage{ulem}

\newcommand{\be}{\begin{equation}}
\newcommand{\ee}{\end{equation}}
\newcommand{\p}{\partial}
\renewcommand{\d}{\partial}
\newcommand{\la}{\label}
\newcommand{\bea}{\begin{eqnarray}}
\newcommand{\eea}{\end{eqnarray}}

\definecolor{cardinal}{rgb}{0.6,0,0}
\definecolor{darkgreen}{rgb}{0,0.4,0}
\definecolor{golden}{rgb}{0.92, 0.7, 0}
\definecolor{midnight}{rgb}{0, 0, 0.5}
\definecolor{darkblue}{rgb}{0, 0, 0.7}

\begin{document}

\title{
	Vortices \& Fractons
	 \vskip 20pt
	 }

\author{Darshil Doshi}
\author{Andrey Gromov}
\affiliation{Brown Theoretical Physics Center \& Department of Physics, Brown University, Providence, Rhode Island 02912, USA}

\date{\today}

\begin{abstract}
We discuss a simple and experimentally available realization of fracton physics.  We note that superfluid vortices form a Hamiltonian system that conserves total dipole moment and trace of the quadrupole moment of vorticity; thereby establishing a relation to a traceless scalar charge theory in two spatial dimensions. Next we consider the limit where the number of vortices is large and show that emergent vortex hydrodynamics also conserves these moments. Finally, we show the motion of vortices and of fractons on curved surfaces agree, thereby opening a route to experimental study of the interplay between fracton physics and curved space. Our conclusions also apply to charged particles in strong magnetic field.
\end{abstract}


\maketitle


\paragraph{Introduction.---} 
Fracton phases of matter are characterized by the presence of immobile or partially mobile local excitations. The constraints on excitation mobility stem from the conservation laws of multipole moments of charge density \cite{pretko2017subdimensional, pretko2017generalized}, \cite{gromov2018towards}.  Phases that support fracton excitations were first discovered in exactly solvable quantum lattice models \cite{chamon2005quantum, haah2011local, vijay2015new}. One systematic approach to characterization and classification of fracton phases is based on tensor \cite{kleinert1982duality, kleinert1983dual, kleinert1983double, xu2006novel,xu2010emergent,rasmussen2016stable, pretko2017subdimensional, pretko2017generalized} and multipole gauge theories (MGT) \cite{gromov2018towards, bulmash2018generalized}.  Recent years have witnessed a significant interest in development and classification of phases of quantum matter supporting fracton excitations \cite{prem2017emergent, prem2018cage, prem2018pinch, slagle2018symmetric, 2019Song, slagle2017fracton, slagle2017quantum, slagleXcube2017, shirley2018foliated, slagle2018foliated, pretko2017emergent, pretko2017higher, devakul2018correlation, devakul2018fractal, you2018subsystem, you2018symmetric, you2019fractonic, weinstein2018absence, wang2019higher, aasen2020topological, ma2017fracton, ma2018topological, yan2018fracton, yan2019rank, yan2019hyperbolic,  schmitz2018recoverable, ma2018higher, gromov2020fracton}, with possible applications ranging from quantum memory to quantum elasticity and quantum gravity. For recent reviews see  \cite{nandkishore2018fractons,pretko2020fracton}. Despite substantial theoretical progress there are few proposals for experimental realization of the fracton physics \cite{yan2019rank, sous2019fractons}.  

 One prominent, but down-to-earth example of excitations with restricted mobility is crystalline defects \cite{pretko2018fracton, gromov2019chiral, radzihovsky2020fractons, gromov2019duality, pretko2018symmetry, kumar2018symmetry, pretko2019crystal, gromov2020duality}. There, dislocations satisfy the \emph{glide constraint} that forces them to move along their Burgers vector, while disclinations are immobile.

In this Note we point out that fracton physics is exhibited by superfluid vortices that have been experimentally observed for many decades. We show that vortices in two spatial dimensions share the mobility constraints with the traceless scalar charge theory (TSCT). We review the Hamiltonian formulation of the vortex dynamics and show that it manifestly conserves dipole and (trace of) quadrupole moments of vorticity. In superfluids, the  vorticity of individual vortices  is quantized and locally conserved, which leads to identification of vorticity with the scalar charge. These conservation laws imply that isolated vortices are immobile, while vortex dipoles move perpendicular to their dipole moment. Both vortices and their dipoles  can be readily created and studied experimentally in superfluid He \cite{donnelly1991quantized}, BECs \cite{neely2010observation, freilich2010real}, polariton superfluids \cite{sanvitto2011all, nardin2011hydrodynamic} and non-linear media \cite{mamaev1996vortex}. We then consider a hydrodynamic limit where the number of vortices becomes large; and collective, hydrodynamic description is applied to the vortices themselves. Remarkably, the resulting hydrodynamics admits a Hamiltonian formulation with Poisson brackets realizing the classical $w_\infty$ algebra. We show that vortex hydrodynamics is also equivalent to scalar charge theory and provide a microscopic collective field theory expression for the rank-$2$ symmetric current. Finally, we discuss the behavior of vortices and fractons on curved manifolds, which can be realized as curved $^4$He films.

\paragraph{Vortices.---} 

We consider a two dimensional incompressible ideal fluid. It is described by the Euler equations
\be\la{eq:Euler}
(\p_0 + u_i\p_i) u_j = - \p_j P\,,
\ee
where $P$ is the pressure and $u_i$ is the velocity field. The combination $\p_0 + u_i\p_i$ is known as material derivative. 
The incompressibility condition implies that $\p_i u_i =0$. Taking curl of \eqref{eq:Euler} we obtain the Helmholtz equation
\be\la{eq:Helmholtz}
(\p_0 + u_i\p_i) \omega = 0\,,
\ee
where $\omega = \epsilon^{ij} \p_i u_j$ is the vorticity.
Eq.\eqref{eq:Helmholtz} admits solutions where the vorticity is concentrated in a finite number of point vortices. The complex velocity field $u_z = u_1 +i u_2$ takes form
\be\la{eq:holomorphicU}
u_z(z) = -i \sum_{\alpha=1}^N \frac{\gamma_\alpha}{z-z_\alpha(t)}\,, \quad \omega(z) = \sum_{\alpha=1}^N \gamma_\alpha \delta^2(z-z_\alpha(t))\,,
\ee
where $z_\alpha(t) = x^\alpha_1(t) + i x^\alpha_2(t)$ (we will switch between complex and Cartesian coordinates at will) is time-dependent position of the $\alpha$-th vortex and $2\pi\gamma_\alpha$ is its circulation; while $\gamma = |\gamma_\alpha|$ is the vortex strength. We have assumed that vorticity is quantized in the units of $\gamma$, which is the case in the superfluids \cite{donnelly1991quantized}.

Remarkably, the vortex coordinates $x^\alpha_i(t)$ form a Hamiltonian system \cite{lin1941motion} 
\bea\la{eq:VortexH}
&&H = -2\pi\sum_{\alpha<\beta}\gamma_\alpha \gamma_\beta \ln |x^\alpha - x^\beta|\,,
\\ \la{eq:VortexPB}
&&\left\{x^\alpha_1, -2\pi\gamma_\beta x^\beta_2 \right\} = \delta^{\alpha\beta}\,, 
\eea
where $\alpha,\beta=1,2,\ldots,N$ label the vortex strength. We refer the reader to \cite{newton2013n, aref2007point} for an in-depth review of the vortex systems. 

Dynamical system (4)-(5) also describes charged particles moving in a strong magnetic field, in the limit of infinite cyclotron frequency, or equivalently, on the lowest Landau level. Consequently, all our results apply verbatim to plasma in strong magnetic field.

In dealing with \eqref{eq:VortexH}-\eqref{eq:VortexPB} it is useful to use the complex coordinates $z_\alpha $. In complex notations the only non-trivial Poisson brackets take the form
\be\la{eq:VortexPBc}
\left\{z_\alpha, \bar{z}_\beta \right\} = i (\pi\gamma_\alpha)^{-1} \delta_{\alpha\beta}\,.
\ee
The equations of motion are
\be\la{eq:eomz}
\dot{\bar z}_\alpha = -i \sum_{\beta=1, \beta\neq \alpha}^N \frac{\gamma_\beta}{z_\alpha - z_\beta}\,.
\ee
It is worth emphasizing that $H$ is not just the potential energy. Due to the non-trivial commutations relations between $z_\alpha$ and $\bar z_\alpha$, $H$ can be viewed as kinetic energy. 

\paragraph{Conservation laws.---} 

Hamiltonian $H$ is translation and rotation invariant. The corresponding integrals of motion are known as \footnote{These are also known as center of circulation and moment of circulation.} \emph{impulse}, $P_i$ and \emph{angular impulse}, $L$ \cite{saffman1992vortex, aref2007point}. They are given by
\be\la{eq:Charges}
P_i = -2\pi\epsilon_{ij} \sum_\alpha \gamma_\alpha x^\alpha_j\,, \qquad L = 2\pi\sum_\alpha \gamma_\alpha x^\alpha_i x^\alpha_j \delta_{ij}\,.
\ee

We recognize in Eqs.\,\eqref{eq:Charges} that impulse is related to the dipole moment of vorticity (also known as center of circulation), while angular impulse corresponds the trace of the quadrupole moment of vorticity, $Q_{ij}$ (also known as moment of circulation), according to
\be
P_i = -\epsilon_{ij} D_j\,, \qquad L = \delta_{ij} Q_{ij}\,.
\ee
Together, the quantities $P_i, L, D_i, Q_{ij}$ form a multipole algebra \footnote{It is important to note that unlike in the case of theories studied in \cite{gromov2018towards}, there are no additional internal symmetries responsible for the conservation of dipole and quadrupole tensors. These conservations, instead, stem from spatial symmetries and non-commutativity.}
\bea
\begin{aligned}
&\left\{L, P_i\right\} = 2\epsilon_{ij} P_j\,, &\left\{P_i, D_j\right\}= -\delta_{ij} \Gamma\,, \\
&\left\{L, D_i\right\} = -2\epsilon_{ij} D_j\,, &\left\{P_i, \delta_{jk}Q_{jk}\right\} = -2 D_i\,,
\end{aligned}
\eea
where we have introduced the total vortex strength
\be
\Gamma = \sum_{\alpha=1}^N\gamma_\alpha\,.
\ee
Thus, \emph{the vortices are equivalent to a traceless scalar charge theory}; where total charge, dipole, and trace of the quadrupole are conserved \cite{pretko2017subdimensional}. Isolated charges are immobile; while isolated dipoles move perpendicular to their dipole moment. Vortices and vortex dipoles are experimentally available with the present day technology.

\paragraph{Mobility constraints.---} 

\begin{figure}
  \includegraphics[width=\linewidth]{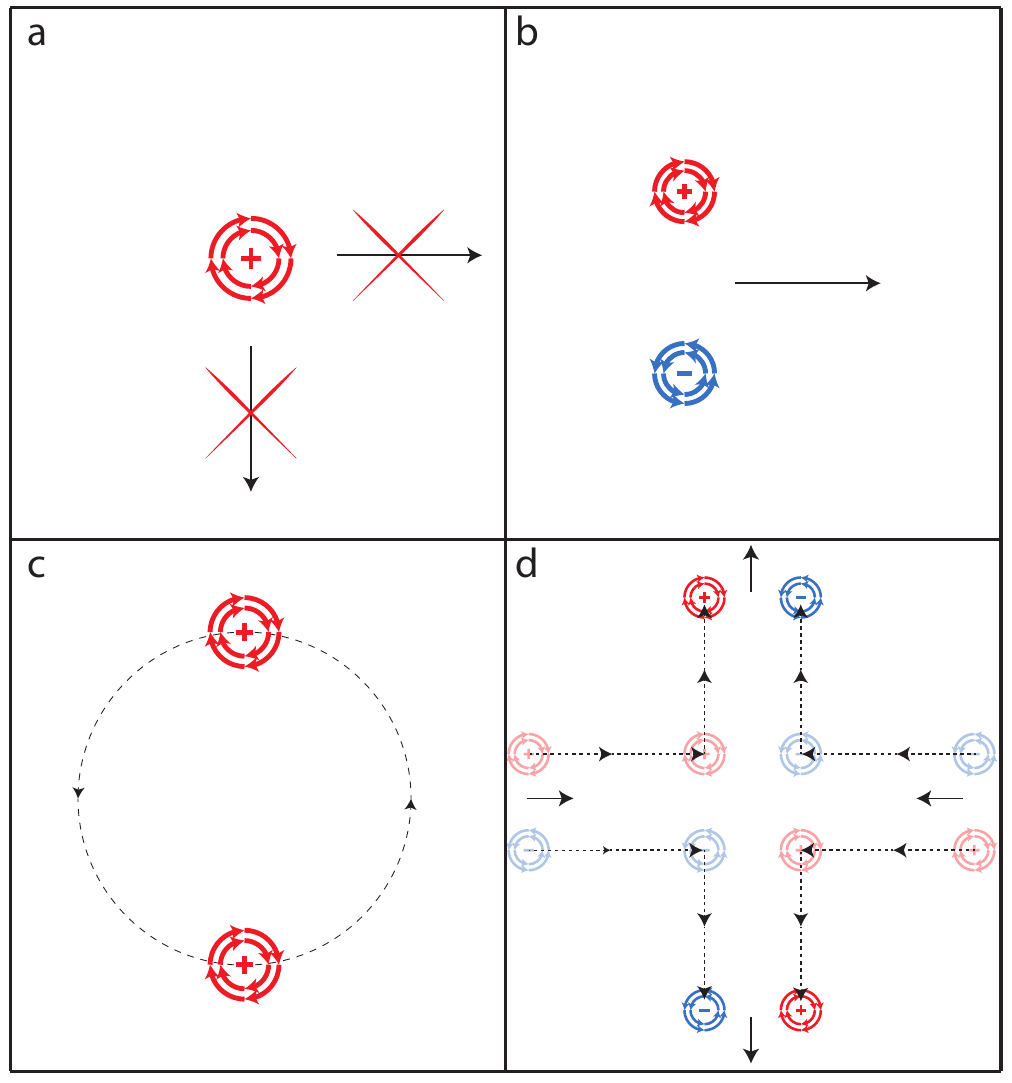}
  \caption{\textbf{a}. An isolated vortex is immobile and corresponds to a fracton. \textbf{b}. Neutral dipole moves perpendicular to its dipole moment -- it is a ``lineon''. \textbf{c}. Charge $2$-dipole moves around center of vorticity. In fractonic context this motion is also possible, albeit never discussed: a pair of identical charges can rotate by constantly emitting dipoles. \textbf{d}. Scattering of two dipoles of opposite dipole moment. After scattering dipole makes a $\pi/2$ turn. }
  \la{fig:vortex}
\end{figure}

Conservation laws \eqref{eq:Charges} imply that motion of many vortices is constrained to preserve the dipole and quadrupole moment. Moreover, since the conserved quantities $(H, D_iD^i, \delta^{ij}Q_{ij})$ are in involution, the problem of $N$ vortices is integrable for $N \leq 3$\footnote{The problem of $4$ vortices with vanishing charge and dipole moment is integrable}. Other typical cases are chaotic \cite{aref2007point}. We discuss the ``fractonic'' motion of vortices next. 

 A single or well-isolated, vortex is immobile. Analogously to fractons, the mass of an isolated vortex is not well-defined. A broad class of definitions \cite{thouless2007vortex} leads to the diverging mass, which agrees with fracton ideas \footnote{We note that there are other ways to define vortex mass that leads to vanishing \cite{baym1983hydrodynamics} or to finite result \cite{simula2018vortex}. This definition results in an answer given in terms of kelvon excitation energy.}. 

Dipole consisting of two vortices with opposite vorticities moves in a straight line perpendicular to its dipole moment. At low temperatures vorticity neutral systems ``condense'' into a gas of neutral dipoles \cite{kraichnan1980two}. The dipole of two vortices of the same vorticities moves in a closed orbit around their ``center of vorticity", while keeping the distance between the two vortices constant. Motion of dipole is illustrated in Fig. \ref{fig:vortex}. Relative distances can only change if the number of vortices is $N\geq 3$ \cite{aref2007point}. 

The quadrupole of two vortex-dipoles exhibits a variety of complex dynamics. One common type of interactions (particularly at low temperature) is scattering between two dipoles as shown in Fig. \ref{fig:vortex}. As a result of scattering a vortex dipole makes a $\pi/2$ turn, which agrees with phenomenology of TSCT.

\paragraph{Statistical mechanics.---} 

Although many-vortex dynamics is chaotic, for certain vortex configurations the \emph{relative} positions of vortices are completely frozen. Such configurations are called vortex crystals or vortex equilibria \cite{aref2002vortex}. The examples include $N$ \emph{identical} collinear vortices situated in the roots of $N$-th Hermite polynomial as well as Adler-Moser polynomials, identical vortices located in the vertices of a regular $N$-polygon, etc. There are many other examples (see \cite{aref2002vortex} for a review). Vortex crystals can move as rigid objects in which case they are referred to as relative equilibria, or can be stationary. Such configurations explore a very small fraction of the phase space. This is immediately obvious since for a vortex system phase space coincides with the configuration space. Vortex crystals emerge experimentally after relaxation of highly turbulent two dimensional flows \cite{fine1995relaxation, schecter1999vortex}.

 It is tempting to compare vortex crystals to the Hilbert space fragmentation seen in quantum dipole conserving systems \cite{pai2019localization, khemani2019local, 2020PollmannFragmentation}. There, Hilbert space ``shatters'' into many disconnected subspaces, within each such subspace either integrability or thermalization is possible.

Mobility constraints combined with the phase space reduction lead to an exotic statistical mechanics of vortices \cite{onsager1949statistical, lundgren1977statistical}. In particular, above certain critical energies vortices experience ``negative temperature'' \cite{onsager1949statistical, montgomery1974statistical}, which follows from the structure of the phase space \footnote{Namely, from the fact that the phase space is finite}.  At negative temperature the vortices of the same vorticity tend to clamp together, which nicely corresponds to gravitational attraction of fractons discussed in \cite{pretko2017emergent}. Vortex crystals may be an obstruction to ergodicity: Clusters of vortices take a very long time to merge \cite{lundgren1977statistical}. To the best of our knowledge, the ergodicity of vortex system is still an open problem \cite{eyink2006onsager}.

\paragraph{Vortex hydrodynamics.---} 

Next we would like to consider a limit where the number of vortices is very large. Due to chaotic behavior and strong interactions between the vortices, this limit admits a description in terms of an emergent hydrodynamics \cite{wiegmann2014anomalous}. We will show that in hydrodynamic limit the dipole and trace of the quadrupole moments are conserved. These conservation laws will be made manifest by re-writing the continuity equation in the rank-$2$ form
\be\la{eq:TSCT}
\dot{\rho} + \p_i \p_j J_{ij} = 0\,,\qquad \text{Tr}(J_{ij})=0\,,
\ee
where $J^{ij}$ is the symmetric, traceless rank-$2$ tensor. Here $\rho = {(2\pi\gamma)}^{-1}\omega$ is the local vortex density. This form of continuity equation implies that the dipole moment and the trace of the quadrupole moment are conserved
\bea
&&\p_0 D_k =\p_0 \int x_k \rho = \int x_k  \p_i \p_j J_{ij}  = 0\,,
\\
&&\p_0 \text{Tr}(Q_{ij}) =\p_0 \int x^2 \rho  = \int \text{Tr}(J_{ij}) =0\,.
\eea
We would like emphasize one subtle difference between traditional TSCT and vortices: The former is non-chiral, while the latter is chiral. In TSCT a dipole moves perpendicular to its dipole moment; while for a vortex dipole, the dipole moment and the direction of motion form a right pair.

Vortex hydrodynamics for the \emph{chiral} flow (\emph{i.e.} when all vortices are of the same vorticity, $\gamma_\alpha=\gamma$) was derived by Wiegmann-Abanov in \cite{wiegmann2014anomalous}. The continuum limit of the vortex Hamiltonian \eqref{eq:VortexH} is \footnote{We assume that the argument of $\ln{\rho}$ is multiplied by an appropriate power of $\ell$. $\ell$ is the length-scale of the patches that we average over when taking the hydrodynamic limit.}
\be \la{eq:AWHam}
H_{\rm WA} = \frac{1}{2} \int \left[v^2 - \eta^2 (\p_i \ln \rho) (\p_i \ln \rho)\right] d^2 r\,,
\ee
where $v_i$ is the \emph{vortex} velocity and $\eta = \frac{\gamma^2}{4}$. Vortex fluid is incompressible $\p_i v_i =0$ and $v_i$ is completely determined by the density through \cite{wiegmann2014anomalous}
\be\la{eq:density-velocity}
\epsilon_{ij} \p_iv_j = 2\pi\gamma\rho + \eta\Delta\ln \rho\,. 
\ee
The Poisson brackets form the classical $w_\infty$ algebra
\be\la{eq:densityalgebra}
\left\{ \rho(x),\rho(x^\prime)\right\} =\epsilon_{rs}\p^\prime_{r}\p_s \left[ \rho(x) \delta(x-x^\prime)\right] \,,
\ee
where $\p^\prime_i = \frac{\p}{\p x_i^\prime}$. Brackets between velocity and density are deduced from \eqref{eq:density-velocity} 
\bea\nonumber
\left\{ v_k(x), \rho(x^\prime) \right\} &&= - \p^\prime_k \left(\rho(x)\delta(x-x^\prime)\right) 
\\
&&- \eta \epsilon_{kj} \p_j \left[ \frac{1}{\rho} \epsilon_{rs}\p^\prime_r\p_s\left(\rho(x) \delta(x-x^\prime)\right) \right].
\eea

We are interested in computing the equation of motion for the density $\rho$
\be
\dot\rho(x) = \left\{H_{\rm WA},\rho(x)\right\}\,.
\ee
Direct calculation gives the continuity equation
\be\la{eq:contrho}
\dot{\rho} + \p_k j_k = 0\quad \Leftrightarrow \quad \dot{\rho} + v_k\p_k \rho = D_0 \rho=0\,,
\ee
where $j_k = \rho v_k$. This is consistent with Helmholtz equation \eqref{eq:Helmholtz}. The consistency is non-trivial since \eqref{eq:Helmholtz} includes the material derivative with $u_i$, while  the material derivative contains $v_i$ in \eqref{eq:contrho}. The equivalence of \eqref{eq:Helmholtz} and \eqref{eq:contrho} is established using  the relation between $u_i$ and $v_i$ \cite{wiegmann2014anomalous}
\be\la{eq:uv}
v_i = u_i - \frac{\gamma}{4} \epsilon_{ij}\p_j \ln \rho\,.
\ee

Using the identity 
\be
\la{eq:Identity}
2\pi\gamma u_i \rho = \epsilon_{ik}\left[\p_j(u_ju_k) - \frac{1}{2}\p_k(u_ju_j)\right]\
\ee
with either \eqref{eq:Helmholtz}  or  \eqref{eq:contrho} we find
\be\la{eq:Jchiral}
j_i =  \p_j J_{ij}\,, \qquad
J_{ij} = \frac{1}{2\pi\gamma}\left(\epsilon_{ik}u_ju_k - \frac{1}{2}\epsilon_{ij}u^2\right) - \frac{\gamma}{4}\epsilon_{ij}\rho\,.
\ee
The anti-symmetric part of $J_{ij}$ drops out from \eqref{eq:TSCT}. In the chiral case an equivalent relation was derived in \cite{bogatskiy2019vortex}.

Emergent hydrodynamics for vortices of positive \emph{and} negative vorticity was developed by Yu-Bradley \cite{yu2017emergent}. The conservation of the impulse and angular impulse holds in their model as well \footnote{Yu-Bradley hydrodynamics involves two fluids and, consequently, two densities: number and charge density. The dipole moment of the charge density is conserved, while the dipole moment of the number density is not}. We will discuss an independent collective field theory derivation of the rank-$2$ conservation law \eqref{eq:TSCT}  for arbitrary number of vortices next.

\paragraph{Collective field theory of vortices.---}
We now turn to the collective form of \eqref{eq:eomz}. Presence of  positive \emph{and} negative vortices is assumed.

Density and current fields are defined as follows 
\be
\begin{aligned}
\rho(z) &= \frac{1}{\gamma}\sum_\alpha \gamma_\alpha\delta(z-z_\alpha)\,,
\\j_z(z) &= \rho(z) v(z) = \frac{1}{\gamma}\sum_\alpha\gamma_\alpha\dot{\bar z}_\alpha\delta(z-z_\alpha)\,.
\end{aligned}
\ee
We will need the complex notation $j_z = j_1 - i j_2$ and the $\delta$-function identity 
\be
\p_z \frac{1}{\bar z} =\p_{\bar z} \frac{1}{z} =\pi \delta(z)\,.
\ee
Time derivative of the density is given by
\be
\dot{\rho} = -\frac{1}{\gamma}\sum_{\alpha=1}^N\Big[    \gamma_\alpha\dot{z}_\alpha \p_{z}\delta(z-z_\alpha)  +  \gamma_\alpha\dot{\bar z}_\alpha \p_{\bar z} \delta(z-z_\alpha)\Big]\,.
\ee
Using \eqref{eq:eomz} this is transformed into the 
\be
\dot{\rho} + \p_z \p_z J_{\bar z \bar z} + \p_{\bar z}\p_{\bar z} J_{zz} = 0\,,
\ee
where we have introduced a \emph{traceless} symmetric tensor current
\be\la{eq:collectiveJ}
J_{zz} = \frac{1}{2\pi i \gamma} \left( \left(\sum_\alpha \frac{ \gamma_\alpha}{z-z_\alpha}\right)^2 + \p_z\sum_\alpha \frac{\gamma_\alpha^2}{z-z_\alpha} \right)\,,
\ee
and $J_{\bar z \bar z} = \bar J_{zz}$. It is crucial that in \eqref{eq:collectiveJ} the second order poles cancel. In Cartesian components the symmetric tensor current is given by
\bea
\la{eq:TensorCurrent}
J_{ij} &=& \frac{1}{2\pi\gamma}\left(\epsilon_{ik}u_ju_k - \frac{1}{2}\epsilon_{ij}u^2\right) - \frac{\gamma}{4}\epsilon_{ij}n\,,
\\
\la{eq:NumberDensity}
n &=& (2\pi \gamma)^{-1}\sum_\alpha |\gamma_\alpha|\delta(z-z_\alpha)\,,
\eea
where we introduced the vortex number density  $n(z)$. This is the central result of the present work: The continuity equation takes form \eqref{eq:TSCT}. The above derivation is general and applies to hydro with vortices of both kinds present. In particular, it applies to the case when total vorticity is $0$.

\paragraph{Curved space.---} 
Symmetric tensor gauge theories do not remain gauge invariant on a curved space \cite{gromov2019chiral}. Furthermore, the conservation law of dipole moment cannot remain unchanged on a curved space. Below, we show that, on a curved space, the dynamics of vortices and the mobility constraints change. Vortices on a curved space have been studied in \cite{hally1980vortex, kimura1999vortex, hally1980stability} and can be experimentally realized in thin $^4\text{He}$ films. Vortex hydrodynamics of chiral flows was generalized to curved spaces in \cite{bogatskiy2019vortex}. Vortex problem on a surface of a sphere is also relevant for geophysical and atmospheric applications. The Helmholtz equation on a curved surface takes form \cite{bogatskiy2019vortex}
\be\la{eq:CurvedHelm}
\dot{\rho} + u_i \nabla_i \left( \rho +  \frac{s}{4\pi} R \right)=0\,,
\ee
where $\nabla_i$ is a covariant derivative, $R$ is Ricci curvature and $s-\frac{1}{2}$ is the geometric spin of a vortex.  Eq.\eqref{eq:CurvedHelm} also takes form \eqref{eq:TSCT} with slightly modified $J_{ij}$ \cite{bogatskiy2019vortex}
\be\la{eq:JR}
J_{ij}(R) = J_{ij}\Big|_{R=0} + \left[\frac{1}{2\pi \gamma} \nabla_i u_j + \frac{\gamma}{2} \epsilon_{ij}\left(\rho - \frac{s}{4\pi} R\right)\right]\,. 
\ee
Note that the last term in \eqref{eq:JR} contributes to the equations of motion only when curvature is inhomogeneous.
We can draw the following conclusion from \eqref{eq:CurvedHelm}-\eqref{eq:JR}. On a surface of constant curvature an isolated vortex remains immobile \cite{kimura1999vortex, dritschel2015motion}, which is consistent with \cite{gromov2019chiral}. A dipole moves along a geodesic that is perpendicular to the dipole moment; which is consistent with the corresponding fracton observations made in \cite{slagle2019symmetric}. 

On a surface of variable curvature an isolated vortex does move: the dipole conservation law is broken and fractonic property is lost; in agreement with \cite{gromov2019chiral}. The potential force acting on an isolated vortex is obtained by differentiating the Robin function \cite{boatto2008vortices}. The dipole moves along a geodesic in the general case \cite{koiller2009vortex}.

\paragraph{Conclusions.---} 

We have established an equivalence between vortex dynamics in two-dimensional superfluids and traceless scalar charge theory. We have shown that vortices provide a Hamiltonian realization of fracton dynamics for any finite number of vortices and in hydrodynamic limit. Thus superfluid vortices provide a readily available platform for experimental realization of fracton quasiparticles.

Similar conservation laws hold in 3 dimensions for vortex lines. We leave the exploration of higher dimensional case, discussion of more refined probes of fracton dynamics in superfluids and BECs such as role of the trap and finite lifetime, generalization to chiral superfluids such as $^3$He and many other open question to future work. Theory of vortices plays central role in statistical approach to turbulence \cite{onsager1949statistical}; where the questions of ergodicity and validity of statistical mechanics are central \cite{eyink2006onsager}. It would be very interesting to see if fracton-inspired ideas can lead to new insight into quantum and classical turbulence as well as the problem of quantization of vortex dynamics. Finally, dynamics of electrons residing in the lowest Landau level is formally identical to that of vortices, consequently we expect applications of fracton inspired ideas to the physics of fractional quantum Hall effect.

\paragraph{Acknowledgments.---}

We thank A. Abanov, A. Bogatskiy, S. Moroz for comments on the manuscript and A. Bogatskiy for bringing the results of \cite{bogatskiy2019vortex} to our attention.
 A.G. was supported by the Brown University.


\bibliography{Bibliography}

\onecolumngrid
\pagebreak

\begin{appendix}
\section{Chiral traceless scalar charge theory}
\label{app:TSCT}
In this Appendix we describe the traceless scalar charge theory that satisfies Eq.(12) and its chiral counterpart. The theory involves a single complex scalar field $\Phi$ and is invariant under the following transformations
\be\la{eq:transform}
\Phi^\prime = e^{if(x)} \Phi\,, \qquad f(x) = \lambda +  \lambda_k x_k +  \zeta |x|^2\,,
\ee
where parameters $\lambda, \lambda_k, \zeta$ are arbitrary. The corresponding conservations laws are that of charge, dipole moment and trace of the quadrupole moment.
The Lagrangian invariant under these transformations can be found as follows. First, we define the invariant derivative operators
\be
D_I(\Phi) = \sigma_I^{ij}\left(\p_i \Phi \p_j \Phi - \Phi \p_i \p_j \Phi\right)\,,
\ee
where $\sigma_I$ are the Pauli matrices and $I=1,3$. Under the \eqref{eq:transform} we have
\be
\delta D_I(\Phi) = \sigma_I^{ij}\p_i \p_j f(x) = 0\,.
\ee
The invariant Lagrangian is constructed using these derivatives
\be\la{eq:sct}
\mathcal L = \dot \Phi^\star \dot \Phi + g_1 |D_1(\Phi)|^2 + g_2 |D_2(\Phi)|^2 + g_1^\prime \text{Re}\left[(\Phi^\star)^2D_1(\Phi)\right] + g_2^\prime \text{Re}\left[(\Phi^\star)^2D_2(\Phi) \right] + \mu |\Phi|^2 + \ldots\,,
\ee
where $\ldots$ stands for the higher order terms and the ``interesting'' terms are the first three.
The theory is invariant under $C_4$, but not $SO(2)$ for the generic values of $g_I,g_I^\prime$. It is $SO(2)$ invariant in the case $g_1=g_2$ and $g_1^\prime = g_2^\prime$. 
The global symmetry \eqref{eq:transform} leads to the conservation law
\be
\dot{\rho} + \p_i \p_j J^{ij} = 0 \,, \qquad \text{Tr}(J_{ij})=0\,.
\ee

In two spatial dimensions it is possible to add a parity-breaking term to the Lagrangian \eqref{eq:sct}. Indeed, consider
\be
\mathcal L_{\rm odd} = i \epsilon^{kl}\delta^{ij}  \left(\p_i \Phi^\star \p_k \Phi^\star - \Phi^\star \p_i \p_k \Phi^\star\right) \left(\p_j \Phi \p_l \Phi - \Phi \p_j \p_l \Phi\right)\,,
\ee
where the overall factor of $i$ is required because the term is imaginary. This term breaks parity due to the presence of the $\epsilon$-tensor. This theory is invariant under \eqref{eq:transform}. Invariance under the quadratic symmetry follows from symmetry of the invariant derivative operators
\bea
\delta \mathcal L_{\rm odd} = i \epsilon^{jk} \left(\p_i \Phi^\star \p_j \Phi^\star - \Phi^\star \p_i \p_j \Phi^\star\right) \p_i \p_k f + \text{c.c.} = i \epsilon^{ij}\left(\p_i \Phi^\star \p_j \Phi^\star - \Phi^\star \p_i \p_j \Phi^\star\right) + \text{c.c.}= 0\,.
\eea

\section{Charged particles in the Lowest Landau Level}
Here we show that the dynamics governing point vortices also describes a 2D system of charged particles in the presence of a perpendicular magnetic field, in the lowest Landau level limit. The Lagrangian for such a system is
\be
\mathcal L = \sum_I \frac{m}{2}\dot{x}^I_i\dot{x}^I_i + \frac{B}{2}\sum_I e^I\epsilon_{ij}x^I_i \dot{x}^I_j - U\sum_{i<j}e^Ie^J\ln{\lvert x^I-x^J\rvert} \,.
\ee
The last term denotes the Coulomb interactions between charged particles. Projecting this onto the LLL is equivalent to taking the limit $m\rightarrow 0$. In this limit, the Lagrangian and the corresponding Hamiltonian are
\bea
\la{eq:lllL}
\mathcal L &=& \frac{B}{2}\sum_I e^I\epsilon_{ij}x^I_i \dot{x}^I_j - H \\
\la{lllH}
H &=& \frac{U}{2}\sum_I e^Ie^J\ln{\lvert x^I-x^J\rvert} \,.
\eea
Note that the Hamiltonian is the same as that in Eq.(\ref{eq:VortexH}), with the vortex strengths $\gamma_\alpha$ replaced by the charges $e^I$. The conjugate momenta are given by
\be
p^I_i = \frac{\d\mathcal L}{\d \dot{x}^I_i} = -Be^I\epsilon_{ij}x^I_j \,,
\ee

The dynamics can then be described by the Poisson bracket
\be
\{ x^I_1,-Be^Jx^J_2\} = \delta^{IJ} \,,
\ee
which is the same as the one in Eq.(\ref{eq:VortexPB}); and implies dynamics similar to point vortices.

The first term in the action is invariant under an infinite set of area-preserving diffeomorphisms; but the Coulomb interaction reduces the set of symmetries to (global) translations and rotations. The corresponding Noether charges, namely, total momentum and angular momentum are related to electric dipole moment and trace of quadrupole moment respectively.
\bea
\text{momentum:} \quad P_i &=& -B\sum_I e^I\epsilon_{ij}x^I_j = -B\epsilon_{ij}D_j\\
\text{angular momentum:}\quad L &=& \sum_I \epsilon_{ij}x^I_ip^I_j = B\sum_I\delta_{ij}x^I_ix^I_i = \delta_{ij}Q_{ij}
\eea
We note that these conserved charges are analogous to those obtained for point vortices in Eq.(\ref{eq:Charges}).

\section{Derivation of the symmetric tensor current from Yu-Bradley hydrodynamics}

Here, we derive the traceless and symmetric tensor current described Eq.(\ref{eq:TensorCurrent}) for the collective field theory describing binary vortex fluid. We start with the current 
\be
j = \rho v = \rho u - \frac{i\gamma}{2}\d_z n \,,
\ee
where $n$ in the last term is the number density defined in Eq.(\ref{eq:NumberDensity}). The last term has been derived in \cite{wiegmann2014anomalous}. In Cartesian coordinates, we can see that the current indeed has the form (\ref{eq:Jchiral}).
\bea
j_i = \rho u_i - \frac{\gamma}{4}\epsilon_{ij}\d_j n 
=\d_j \left[ \epsilon_{ik}u_ju_k - \frac{1}{2}\epsilon_{ij}u^2 - \frac{\gamma}{4}\epsilon_{ij} n \right]
\eea
We've used the identity (\ref{eq:Identity}) in the second equation. The quantity inside the square bracket can then be identified with the tensor current $J_{ij}$,
\be
J_{ij} = \epsilon_{ik}u_ju_k - \frac{1}{2}\epsilon_{ij}u^2 - \frac{\gamma}{4}\epsilon_{ij}n \,.
\ee

\end{appendix}
\end{document}